\documentstyle[12pt]{article}
\evensidemargin\oddsidemargin
\begin{document}
\count0 = 1
\begin{titlepage}
\vspace {20mm}
\begin{center}
\  LOCAL REFERENCE FRAMES\\
\smallskip
 AND QUANTUM SPACE-TIME   \\
\vspace{6mm}

\bf{S.N. MAYBUROV}
\vspace{6mm}

\small{Lebedev Institute of Physics}\\
\small{Leninsky pr. 53, Moscow Russia, 117924}\\

\vspace{15mm}
\small{\bf Abstract}
\end {center}
\vspace{3mm}

\small{
 We argue that the account of Reference Frames quantum properties
 must change the standard space-time picture accepted in
 Quantum Mechanics. If RF is connected with some 
macroscopic solid object then its free quantum motion -
wave packet smearing results in 
additional uncertainty into the measurement of test particle coordinate.
It makes incorrect the use of
Galilean or Lorentz space-time transformations between two RF  and
 the special quantum space-time transformations are formulated.
 It results in generalized   Klein- Gordon equation
 which depends on  RF mass.  Both space and time coordinates
become the operators. In particular RF proper time becomes the operator
 depending of momentums spectra of this RF wave packet , from the point of
 view of other observer.} 

\vspace{34mm}
\small {Submitted to Phys. Lett. A }\\
\vspace {10mm}
\small {Talk given at 'Symmetry in Physics' conference\\
 , Dubna , July ,1997}
\vspace{28mm}
\small {---------}\\
\small {  * E-mail  Mayburov@sgi.lpi.msk.su}
\end{titlepage}
\begin{sloppypar}
\section{Introduction}

Some years ago  and Kaufherr have shown that in nonrelativistic
Quantum Mechanics (QM) the correct definition of physical reference 
frame (RF) must differ from commonly accepted one, which in fact was
transferred copiously from Classical Physics $\cite{Aha}$. The main reason  
is that in strict QM framework one should account the quantum properties 
not only of studied object, but also RF, despite the possible practical
 smallness.   
 The most simple  of this properties is the existence of Schrodinger
 wave packet of free macroscopic object with which RF is usually
associated $\cite {Schiff}$. If this is the case it inevitably introduces
additional uncertainty in the measurement of object space coordinate.
Furthermore  this effect account results in the coordinate transformations
between two quantum RFs,  principally different from the Galilean ones
$\cite{Aha,Tol}$.

Further studies of quantum RF effects can help
 to understand the aspects of Quantum space-time which now are
extensively investigated   $\cite{Dop}$. 
 The importance of RF quantum properties was noticed already in 
Quantum Gravity studies  $\cite {Rov,Unr}$. In this paper we discuss mainly
the construction of relativistically covariant  quantum RF theory,
by the analysis of some simple models. It will be shown that the state vector
transformations between two RF obeys to relativistic invariance
principles, but due to dependence on RF state vectors
differs from Poincare Group
transformations. The time ascribed to such RF becomes the operator.
In particular  proper time in each RF
is the operator depending on its momentum, which introduces
 the quantum fluctuations in
the classical Lorentz time boost in moving RF time measurements.

Our paper is organized as follows : in the rest of this chapter
 our model of quantum RF will be formulated and its compatibility with
 Quantum Measurement Theory discussed.
In a chapter 2 the new canonical formalism of quantum RF states and their
transformations  developed, which is more simple and realistic then 
proposed in $\cite{Aha}$.
  The relativistic equations for quantum RF
and the resulting quantum space-time properties are regarded in 
chapter 3 . In a final chapter
the obtained results and their interpretation are discussed.
 
 In QM framework the system regarded as RF presumably should be able
 to measure 
the observables of studied quantum states and due to it
 to  include measuring devices - detectors.
 As the realistic example of such RF we can regard the photoemulsion plate or
the diamond crystal which can measure microparticle position relative to its
c.m. and simultaneously record it.
At first sight it seems that  due to it quantum RF problem must use  
as its basis the detailed model of state vector collapse.
 Yet despite the multiple proposals up to now well established
theory of collapse doesn't exist  $\cite{May,Desp}$.
  Alternatively we'll show that our problem premises doesn't connected
 directly with the state vector collapse mechanism and
 as the result 
 we can use two simple assumptions
about the RF and detector states properties
  which are in the same time rather weak.
  The first one is that  RF consists of 
 finite number of atoms (usually rigidly connected)  and have the finite
 mass.
 Our second assumption needs some preliminary comments.
 It's well known that the solution of Schrodinger equation for
  any free quantum system  consisting of $N$
constituents can be presented as :
\begin {equation} 
  \Psi(\vec{r}_1,...,\vec{r}_n,t)=
\sum c_l\Phi^c_l(\vec{R}_c,t)*\phi_l(\vec{r}_{i,j},t)
\end {equation}
where  center of mass coordinate $\vec{R}_c=\sum m_i*\vec{r}_i/M$.  
 $\vec{r}_{i,j}=\vec{r}_i-\vec{r}_j$  are  the relative or 'internal'
 coordinates of constituents
 $\cite{Schiff}$. Here $\Phi^c_l$ describes the c.m. motion of the system.
 It demonstrates that the evolution of the
 system  is separated into the external
evolution  of  pointlike particle M and the internal evolution
 completely defined by $\phi_l(\vec{r}_{ij},t)$
 So the internal evolution is
 independent of whether the  system is localized
 in the standard macroscopic 'absolute' reference frame (ARF)
 or not. Relativistic QM and Field Theory  evidences that the
 factorization of c.m.
 and relative motion holds true even for nonpotential forces and 
 variable $N$ in the secondarily quantized systems $\cite{Schw}$.
 Moreover this factorization expected to be correct for nonrelativistic
 systems 
 where binding energy is much less then its mass $m_1$, which is
 characteristic for the most of real detectors. Consequently it's reasonable
to extend this result on the detector states despite we don't know
their exact structure. We'll use it quite restrictively and 
  assume that
   the factorization of the c.m. motion holds for RF
  only in the time interval $T$
 from RF preparation moment , until the act of measurement starts
,i.e. when the measured particle collides with it. More exactly our second and
 last assumption about observer properties is    
that  during period $T$ its state is described by  wave function
generalizing  (1) of the form 
$$
 \Psi(R_c,q,t)=\sum c_l\Phi^c_l(R_c,t)*\phi_l(q,t)
$$
 where $q$ denote all internal detector degrees of
freedom which evolve during $T$ according to Schrodinger equation
(or some field equation). Its possible violation at later time when
particle state collapse occurs is unimportant for our model.


 To simplify our calculations
we'll take below all $c_l=0$ except $c_1$ which wouldn't influence our final
 results.
The common opinion is that to observe experimentally 
measurable smearing of macroscopic object
demands too large time , but for some mesascopic experiments
it can be reasonably small to be tested in the laboratory conditions
$\cite{May2}$.  
 We don't consider in our study the influence of RF recoil
 effects on the measurements results which can be made arbitrarily
small $\cite {Aha}$.  


\section{Quantum Coordinates Transformations }
We remind here only briefly the physical meaning of quantum RF, because
in  pioneering paper $\cite{Aha}$ authors analyzed in detail 
gedankenexperiments with quantum RF and interested reader can find
the discussion there.
Suppose that in absolute RF ( in two dimensions $x,y$)
the wave packet of RF $F^1$  is 
  described by wave function $\psi_1(x)\phi_1(y)$. 
 The test particle $n$
  with mass $m_n=m_2$ belongs to   narrow beam which average velocity
is  orthogonal to $x$ axe
 and  its wave function can be presented as $\psi_n(x)\phi_n(y)$.
We want to find $n$ wave function for the observer situated in
 $F^1$ rest frame. Formally it can be done by means of canonical 
transformations described in detail below. But from the qualitative side
it follows that in the  simplest case when $n$ beam is well localized  
and $\psi_n(x)$ can be approximated
 by  delta-function $\delta(x-x_b)$ this wave function in $F^1$
 (its x-component)
$\psi'_n(x_n)=\psi_1(x_n-x_b)$. It shows that if $F^1$ wave packet
have average width $\sigma_x$ then from ths 'point of view'
of observer in $F^1$ each object localized in ARF acquires wave packet of the
same width.   

Let's assume that $F^1$ includes
detector $D_0$ which can measure the distance between $n$
and $F^1$ c.m. 
Then considering the collapse induced by $n,D_0$ interaction
 $F^1$ and ARF observers will treat the same event unambiguously
as  $n$ detection  (or its flight through $D_0$). In
observer reference frame $F^1$  it
reveals itself by the detection and amplification process in $D_o$
 initiated by $n$ absorption. For ARF
the collapse results from the nonobservation of neutron in a due time
 - so called negative result experiment. So 
the signal in $F^1$ will have  the same relative probability as in
ARF. This measurement means not only the reduction of $\psi'_n$ in $F^1$
, but also the reduction of $\psi_1$
in ARF. Due to it we'll assume 
always that all measurements  are performed on
 quantum ensemble of observers $F^1$. It means that each event is resulted 
 from the interaction between the 'fresh' RF and particle
 ,prepared both in the specified quantum
 states ,alike the particle alone in the standard experiment.
 As we 
have no reason to assume that the transition from ARF to $F^1$
 can transfer 
pure states to mixed ones we must conclude that this distribution
is defined by neutron wave function in $F^1$. 
 It means that the result of measurement in $F^1$ is also
described by QM Reduction postulate, i.e. that initial state during
 the measurement by RF detector evolve into the mixture of
the measured observable eigenstates.

After this qualitative example we'll regard
the general nonrelativistic formalism, which 
differ from described in $\cite{Aha}$.
 Consider the system $S_N$ of $N$ objects $B^i$  which
 include $N_g$ pointlike 'particles' $G^i$ and $N_f$ frames $F^i$,
which in principle can have also some internal
 degrees of freedom described by (1).
For the start we'll assume that particles and RF coordinates 
$\vec{r}_i$ are given
 in absolute (classical) ARF having very large
mass $m_A$ and formally 
having coordinate $\vec{r_A}=0$ and momentum $\vec{p}_A=0$.
 At the later stage of the study this assumption can be abandoned and
only relations between quantum RF regarded.
We should find two transformation operators - from ARF to quantum RF
,and between two quantum RF, but it'll be shown that in most general 
approach  they coincide .
We'll use Jacoby canonical
coordinates  $\vec{q}_j^l , 1\leq j\leq N$, which for $F^l$ are equal :
\begin{eqnarray}
 ,\vec{q}^l_{i}=\frac{\sum\limits^N_{j=i+1}m^l_j\vec{r}^l_{j}}{M^{l,i+1}_N}
-\vec{r}^l_{i}
\quad , \quad \vec{q}^l_N=\vec{R}_{cm}-\vec{r}_A \quad \label{B1}
\end{eqnarray}
Here $\vec{r}^1_j=\vec{r}_j , m^1_j=m_j$ , and  for $l>1$ : 
$$
\vec{r}^l_j=\vec{r}_j ,\quad j>l, \quad
\vec{r}^l_j=\vec{r}_{j-1},\quad 1\leq j\leq l ,\quad 
 \vec{r}^l_1=\vec{r}_l
$$
The same relations connect for $m^l_j$ and $m_j$.
  $M^{l,i}_n=\sum\limits^{n}_{j=i}m^l_j$ (if upper index $i,l$ are omitted,
it assumed that $i,l=1$)
 Conjugated to $q^1_i$ $(i=1,N)$ canonical momentums are :
\begin {equation} 
\vec{\pi}^1_i=
\mu'_i(\frac{\vec{p}^s_{i+1}}{M^{i+1}_N}-\frac{\vec{p}_{i}}{m_{i}}) ,\quad
\vec{\pi}^1_N=\vec{p}^s_1 \label{B2}
\end {equation}
where $\vec{p}^s_i=\sum\limits^{N}_{j=i}\vec{p}_j$
,and reduced mass
 $ \mu'^{-1}_i=(M^{i+1}_N)^{-1}+m_{i}^{-1} $ .

The relative coordinates  $\vec{r}_j-\vec{r}_1$ can be represented
 as the linear sum of several coordinates
 $\vec{q}^1_i$ ; they don't
constitute canonical set due to quantum motion of $F^1$.

 We consider first
 the transformation between two quantum RF and start from
the simplest case $N_f=2,N_g=0$. This is just the space reflection
of  $F^1$ coordinate $\vec{q}^2_1=-\vec{q}^1_1$ performed
 by the parity operator $\hat{P}_1$. 
 The next case $N_f=2,N_g=1$ is 
 $\vec{q}^1$ coordinates bilinear transformation exchanging $\vec{r}_2,
\vec{r}_1$ :
\begin {equation}
  q^2_{1,2}=\hat{U}_{2,1}q^1\hat{U}^{+}_{2,1}=a_{1,2}q^1_1+b_{1,2}q^1_2
\label {B3}
\end {equation}
Corresponding unitary operator can be decomposed as 
$\hat{U}=\hat{C_2}\hat{R}\hat{C_1}$ ,where $\hat{C_{1,2}}$
are the dilatation  operators,
 which action changes the coordinate scale. For example $\hat{C}_1$
results in $\vec{q}^1_i=c^i_{1}\vec{q}^1_i$, where
 $c^i_{1,2}$ proportional to $\mu'_i$.

 $\hat{R}$ is the rotation on $\vec{q}^1_{1,2}$ intermediate
coordinates hypersurface on the angle :
$$
\beta=-\arccos[\frac{m_2m_1}{(m_3+m_2)(m_1+m_3)}]^\frac{1}{2}
$$
 For the general case $N>3$ it's possible nonethereless
to decompose the transformation from $F^j$ to $F^k$ 
as the product of analogous bilinear operators. Really if to denote
as $\hat{S}_{i+1,i}$ the operator exchanging $F^i,F^{i+1}$  in $\vec{q}^1$ set,
As follows from (\ref{B1}) it changes in fact only
$\vec{q}^1_i,\vec{q}^1_{i+1}$ pair.  $\hat{U}_{2,1}=\hat{S}_{2,1}$
and all $\hat{S}_{j,j-1}$ have the analogous form , changing only parameters
$\beta,c^i_k$.   
Then the transformation operator
 from $F^1$ to $F^k$ is :
\begin {equation}
   \hat{U}_{k,1}=\hat{S}_{2,1}\hat{S}_{3,2}...\hat{S}_{k,k-1} \label {B4}
\end {equation}
It follows immediately that the transformation from $F^j$ to $F^k$ is
 $\hat{U}_{j,k}=\hat{U}_{k,1}\hat{U}^{-1}_{j,1}$.

To find the transformation operator from the classical ARF
 to $F^1$ we'll regard ARF as the quantum  object $B^{N+1}$
with infinite $m_{N+1}$ belonging to extended system
 $S_{N+1}$. 
ARF 'classical' set is $\vec{q}^A_i=\vec{r}_{i}-\vec{r}_A$ , but acting
by parity operators we'll transform it to $q^A_i=-q^A_i$. Then it's easy
to see that for $S_{N+1}$ $q^1_i=q^A_i$ as follows from (\ref{B1}). 
 Note that formally we can regard also each
particle  $G^j$ as RF and perform for them the transformations
 $\hat{S}_{j+1,j}$ described above. 
Then omitting simple calculations we obtain that operator
performing transformations from ARF to $F^1$ 
 is equal to $\hat{U}_{A,1}=\hat{U}_{N+1,1}$ for infinite $m_{N+1}$.
In this case  new $\vec{q}^l$ set  for $S_{N+1}$ 
 can be rewritten as the function of $\vec{r}^l_i,
\vec{r}^l_{N+1}=\vec{r}^l_A$  of (\ref{B1}) to which formally must be added
 $\vec{q}^{l}_{N+1}=\vec{r}_A-\vec{r}_E$ 
,where $E$ is some other classical RF.

 The free Hamiltonian of the system objects motion in ARF is   :
\begin {equation} 
\hat{H}=\hat{H}_s+\hat{H}_c=\frac{(\vec{\pi}^{1}_N)^2}{2M_N}+
\sum\limits_{j=1}^{N-1}\frac{(\vec{\pi}^{1}_i)^2}{2\mu'_i} \label {B5}
\end {equation}
Hamiltonian of $S_N$ in $F^1$ should depend on relative $B^i$ momentums
only , so we can regard $\hat{H}_c$ as the candidate for its role.
Yet relativistic analysis given below introduces some corrections to
 $\hat{H}_c$.
 Note that even settling
$\vec{r_1}=0$, $\vec{p_1}=0$ for $F^1$ we must account them as the operators in
commutation relations ,as was stressed in $\cite{Dir}$.

 This transformations becomes more complicated if we take into account 
the quantum rotation of our RF relative to ARF, which introduce 
additional angular uncertainty into objects coordinates $\cite{Aha}$.
  We'll  consider here only 
  2-dimensional rotations, for which this uncertainty
is connected with $F^1$ orientation
relative to ARF axes. If $F^1$ is the solid object its orientation  
 relative to ARF can be extracted from the relative (internal) coordinates 
 of $F^1$ constituents (atoms). For the simplicity we assume that
 $F^1$ have the dipole structure and all its mass concentrated around
2 points $\vec{r}_{a1},\vec{r}_{b1}$ so that this $F^1$ internal coordinate
is   $\vec{r}_{a1}-\vec{r}_{b1}$ or in polar coordinates $r^d_{1},\theta_1$.
Note that $r^d_1$ is observable which eigenvalue defined by $F^1$ 
constituents interaction. 
Even if $F^1$ have some complex form its orientation defined by the
same single observable $\theta_1$ and only rotational effective mass $m_1^d$
 will depend on this.  Thus after
 performing coordinate transformation $\hat{U}_{A,1}$ from ARF
to $F^1$ c.m.  we'll rotate all the 
objects (including ARF) around it on the uncertain angle $\theta_1$
,so the complete transformation is
 $\hat{U}^T_{A,1}=\hat{U}^R_{A,1}\hat{U}_{A,1}$.
In its turn this rotation operator can be decomposed as
$\hat{U}^R_{A,1}=\hat{U}^c_{A,1}\hat{U}^d_{A,1}$, representing the
rotation of objects c.m coordinates $\vec{q}^1_i$ and $F^i$
 constituents coordinates.
Their calculations are in fact straightforward and follows directly from the
properties of standard orbital momentum operator so we omit the details.
$\theta_1$is independent of $F^1$ c.m. coordinate $\vec{r}_1$ and due to it
the transformation of $\vec{q}^1_i$ is performed analogously to rotation on
fixed angle.
 Their  transformation operator is: 
\begin {equation}
\hat{U}^c_{A,1}=e^{-i\theta_1L_z} ,\quad
L_z=\sum\limits^{N}_{i=1}l_{zi} ,\quad 
 l_{zi}=-id/d\alpha_i \label{B6}
\end {equation}
 where $\alpha_i$ is the polar angle coordinate of $\vec{q}^1_i$ .
It results in coordinate transformation :
\begin {eqnarray}
  q^{1r}_{xi}=q^1_{xi}\cos\theta_1+q^1_{yi}\sin\theta_1 \label{B7}\\
  q^{1r}_{yi}=-q^1_{xi}\sin\theta_1+q^1_{yi}\cos\theta_1 \nonumber
\end {eqnarray}
So the new polar angle is $\alpha^r_i=\alpha_i-\theta_1$.
The transformation of canonical momentums is analogous :

\begin {eqnarray}
 \pi^{1r}_{xi}=\pi^1_{xi}\cos\theta_1+\pi^1_{yi}\sin\theta_1 \label {B8}\\
 \pi^{1,r}_{yi}=-\pi^1_{xi}\sin\theta_1+\pi^1_{yi}\cos\theta_1 \nonumber
\end {eqnarray}
As can be easily checked Hamiltonian $\hat{H}_c$ of (\ref{B5})
  is invariant under this transformation.

The rotation of $F^1$ is performed by the operator $\hat{U}^d_{A,1}$, 
which action settles $\theta_1$ to zero ,and introduce in place of it
the new observable $\theta^r_1$ which corresponds to ARF  angle in $F^1$. 
 If to denote  $\theta_j$ - orientation angles for $F^j$ constituents and
 $l^d_{j}=-i\frac{\partial}{\partial\theta_j}$ their orbital momentums
, then the transformation operator is :
\begin {equation} \label{B9}
\hat{U}^d_{A,1}=\hat{P}^d_1exp(i\theta_1 L_d) ,
\quad L_d=\sum\limits_{i=2}^{N_f}l^d_{i} 
\end {equation}
It results in new canonical observables :
\begin {eqnarray}
\theta^r_j=\theta_j-\theta_1 ,\quad l'^d_{j}=l^d_{j} \quad j\neq1 \label{B10}\\
\theta^r_1=\theta^r_A=-\theta_1 ,
 \quad l'^d_1=l^d_{A}=-l^d_{1}-L_d \nonumber
\end {eqnarray}
where $\hat{P}^d_1$ - parity operator for $\theta_1$.
The new coordinates can be interpreted as corresponding to 
$F^1$ dipole rest frame , where its own angle $\theta'_1$ is fixed to zero
but ARF angle in $F^1$ $\theta^r_1$ becomes uncertain and formally ARF acquires
the orbital momentum $l^d_{A}$.

Analogous considerations permit to find rotational transformation
$\hat{U}^R_{i,1}$ from $F^1$ to $F^i$. It means the
additional rotation
of all the objects on the  angle $\theta^r_i=\theta_i-\theta_1$. 
Consequently the form of $\hat{U}^c$ operator for it is
 conserved and it's just necessary to substitute  $\theta^r_i$ in it as
parameter.
Operator $\hat{U}^d_{i,1}$ can be expressed as :
\begin {equation}
\hat{U}^d_{i,1}=\hat{U}^d_{A,i}(\hat{U}^d_{A,1})^{-1} \label {B11}
\end {equation}
where $\hat{U}^d_{A,i}$ can be easily found from (\ref{B9}).

 Assuming that any $F^i$ have analogous to $F^1$ dipole form
 the part of constituents relative motion Hamiltonian which depends
on orientation in ARF :
\begin {equation} 
\label {B12} 
 \hat{H}_i=\sum\limits_{j=1}^{N_f}\frac{1}{m^d_jr^{d2}_{j}}
\frac{\partial^2}{\partial\theta_j^2}
\end {equation}
where $m^d_j$ is the effective mass of the rotational moment which
for the dipole is equal to its reduced mass. It follows that this operator
is  invariant of $\hat{U}^d_{A,1}$ transformations.



So we get the conclusive and noncontroversial description of
$G^i$ and $F^i$ coordinate transformation to $F^1$ defined by Hamiltonian
 $\hat{H}_c+\hat{H}_i$ .
 Yet this transformation can result in the change of the objects $G^i$
 wave functions $\Psi^1$ in $F^1$ which will depend on $F^1$ orbital
 momentum
which should be accounted performing the initial functions  transformation.

For $d=3$ the mathematical calculations are analogous , but more tedious ,
if to account that any rotation in space can be decomposed as
three consequent rotations in the specified mutually orthogonal planes. 
  We omit this calculations here, and just explain 
 new features appearing. To describe
  this rotation $F^1$ should have the necessary geometric structure
, the simplest of which is the  triangle $abc$ with masses
concentrated in its vertexes. Then $Z'$ axe can be chosen to be orthogonal 
to the triangle plane and $X'$ directed along $ab$ side.
 Then the transformation
which aligns ARF and $F^1$ axes can be performed  rotating consequently
ARF around $X',Y',Z'$ on the uncertain angles $\theta_x,\theta_y
, \theta_z$. Each of three operators performing it is the analog of
 $\hat{U}^R_{A,1}$ described above.

Jacoby formalism described here is more simple then the formalism of
$\cite{Aha}$ and can be easily developed for relativistic quantum RF
description.    

\section  {Relativistic Equations}
The relativistic covariant formalism of quantum RF will be studied with the
 model of  relativistic wave packets
of macroscopic objects  regarded as quantum RF .  
 It supposed that in RF 
  all its constituents spins and orbital momentums are roughly compensated
and can be neglected. For the simplicity we'll
 neglect quantum rotation effects, described in the former chapter.

In nonrelativistic mechanics time $t$ is universal and 
 is independent of observer. In relativistic
case  each observer in principle has its own proper time $\tau$ 
measured by his clocks, which is
presented in evolution equation  for any object in this RF .
We don't know yet the origin of the physical time , but phenomenologically 
 we can associate it with the clock hands motion,
or more exactly with the measurement and recording of their current position
 by observer $\cite{Per}$. 
This motion is stipulated by some irreversible processes
which are practically unstudied on quantum level. It seems
there is a strong and deep analogy between irreversible wave function
collapse in the measurement and clock hands motion+measurement
 which can be regarded as
the system self-measurement $\cite{Hor}$.
Meanwhile without choosing one or other mechanism , it's possible
 to assume that as in the case of the position measurement this internal
 processes can be disentangled from the clocks c.m. motion.
 Then clocks $+$ observer
  $F^2$ wave packet evolution can be described by the
 relativistic equation 
  for their c.m. motion relative to RF $F^1$ and $F^2$ internal degrees of
freedom evolution which define its proper time $\tau_2$ supposedly are
factorized from it.
 In this packet
 different momentums and consequently velocities  relative to
external observer $F^1$ are presented. It makes impossible to connect
external time $\tau_1$ and $F^2$ proper time $\tau_2$ by any Lorentz
transformation, characterized by unique definite
Lorentz factor ${\gamma}(\vec{v})$ $\cite{Lan}$.

To illustrate the main idea we remind the well-known
situation with the relativistic lifetime dilation of unstable particles
or metastable atoms. Imagine that  the prepared beam of them 
is the superposition of two or more momentums eigenstates having
different Lorentz factors $\gamma_i$.
 Then detecting  their  decay  products  we'll
find the superposition of several lifetime exponents , resulting from the
fact that for each beam component Lorentz time boost has its own value.
If in some sense this unstable state can be regarded as elementary
 clock when their time rate for the external observer is defined by the
superposition of Lorentz boosts responding to this momentums.

This arguments suppose that the proper time of any quantum RF
 being the parameter in his rest frame simultaneously
 will be the operator for  other quantum RF.
If this is the case the proper time $\tau_2$ of $F^2$ in $F^1$ can be
 the parameter
depending operator, where parameter is $\tau_1$.
$$ 
    \hat{\tau}_2=\hat{F}(\tau_1)=
 \hat{B}_{12}\tau_1
$$
 ,where $\hat{B}_{12}$ can be called Lorentz boost operator,which 
can be the function of $F^1,F^2$ relative momentum. We regard $F^1$ proper
time as parameter, which means that the possible quantum fluctuations
of $F^1$ clocks are supposed to be small $\cite{Per}$.

 To define $\hat{B}_{12}$
it's necessary  to find $F^2$ Hamiltonian for
which we consider first the relativistic motion of free spinless particle
 $G^2$ in $F^1$. 
Obviously in relativistic case Hamiltonian of relative motion
 of very heavy RF and light particle
  should approximate Klein-Gordon square root
 Hamiltonian, but in general it can differ from it $\cite{Schw}$.
The main idea  how to find it is the same as in nonrelativistic case
 : to separate the system c. m.
 motion and the relative motion of the system parts  $\cite{Coe}$.

 We'll consider first the evolution of system $S_2$ of
 RF $F^1$ and the particle $G^2$
which momentums $\vec{p}_i$ are defined in some classical ARF.
If to regard  initially prepared states including only
positive energy components (not considering antiparticles at this stage)
 , then their joint state vector evolution in 
ARF is described by square root Hamiltonian $\cite{Schw}$:
\begin {equation}
-i\frac{d \psi^0}{d\tau_0}=[(m_1^2+\vec{p}_{1}^2)^{\frac{1}{2}}+
(m_2^2+\vec{p}_{2}^2)^{\frac{1}{2}}]\psi^0 \label {C0}
\end {equation}
in momentum representation.
 To obtain from it the Hamiltonian of $G^2$ in $F^1$ we remind that the
objects relative motion is characterized by their invariant mass  $s^m$,
which is equal to system total energy in its c.m.s..
 In our case it's equal to :
\begin {equation}
s^m_{12}=(m_1^2+\vec{q}_{12}^2)^{\frac{1}{2}}+
(m_2^2+\vec{q}_{12}^2)^{\frac{1}{2}} \label {C1}
\end {equation}
,where $\vec{q}_{12}$ is  $G^2$  momentum in c.m.s. 
 If to define
$S_2$ total momentum in  its c.m.s. $p^s_{\mu}$, which 4-th component is 
 $s^m_{12}$, then transforming it  to $F^1$ rest frame one obtains 
4-vector $p^1_{\mu}$ which 4-component is equal :  
\begin {equation}
  E^1=[(s^{m}_{12})^2+\vec{p}^2_{12}]^\frac{1}{2}=
m_1+(m_2^2+\vec{p}^2_{12})^\frac{1}{2} \label {C2}
\end {equation}  
Here $\vec{p}_{12}$ is classical $G^2$ momentum in
$F^1$ rest frame
\begin {equation}
  \vec{p}_{12}=\frac{s^m_{12}\vec{q}_{12}}{m_1}=\frac{E_1\vec{p}_2
-E_2\vec{p}_1}{m_1}  \label {C3} 
\end {equation}
where $E_{1},E_{2}$ are the energies
  of $F^1,G^2$ in ARF.
 From this classical calculations and Correspondence
principle (which in relativistic QM should be applied carefully)
 we can  regard  $E^1$ as possible form of Hamiltonian $\hat{H}^1$
in $F^1$ rest frame
and the evolution equation for $G_2$ 
 for corresponding proper time $\tau_1$ is :
\begin {equation} 
\hat{H}^1\psi^1(\vec{p}_{12},\tau_1)
 =-i\frac{d\psi^1(\vec{p}_{12},\tau_1)}{d\tau_1} \label {C4}
\end {equation}
It's easy to note that $\hat{H}^1$ depends only of relative motion
observables and in particular can be rewritten as function of $\vec{q}_{12}$.
 This equation will coincide with Klein-Gordon one,
 if we 
consider  $m_1$ as arbitrary constant added to $G^2$ energy. 
Consequently we can use in $F^1$ the same momentum eigenstates spectral 
decomposition and the states scalar product $\cite{Schw}$.

 Space coordinate operator in $F^1$ is difficult
 to define unambiguously, as usual in relativistic QM, but
  Newton-Wigner ansatz can be used without complications $\cite{Wig}$.
\begin {equation}
 x_{12}=i\frac{d}{dp_{x,12}}-i\frac{p_{x,12}}{2(E^1-m_1)^2} \label {C5}
\end {equation}
 
In this framework $F^2$ proper time operator
 $\hat{\tau}_2$ in $F^1$ can be found  
 from the correspondence with the classical Lorentz time boost
appearing in moving clock time measurement
 $\Delta t_2=\gamma_2^{-1}\Delta t_1$, where $\gamma_2$ is $F^2$ Lorentz factor
$\cite {Lan}$. 
Rewriting it as function of  momentums in $F^1$ one obtains :
\begin {equation} 
    \hat{\tau}_2
= (\vec{p}_{12}^2+m_2^2)^{-\frac{1}{2}}m_2\tau_1  \label {C6}
\end {equation}
Due to its  dependence only on  $\tau_1$ - parameter and momentums
this operator is self-adjoint and doesn't suppose
the use of POV measures, used in some models for time operator
 $\cite{Bus}$.
Note that the operators $\hat{x}_{12},\hat{\tau}_2$ doesn't 
commute , due to $\hat{\tau}_2$ dependence on $\hat{p}_{x,12}$.
This result is analogous to  Noncommutative Geometry
and Quantum Groups predictions for quantum space-time at Plank scale
, yet our scale is much larger $\cite{Dop}$.
 Note that this approach is completely symmetrical and the
 operator analogous to (\ref{C6}) relates the
 time $\hat{\tau_1}$ in $F^1$ and $F^2$ proper time
- parameter $\tau_2$.
By himself (or itself ) $F^2$ can't find any consequences of time arrows
superposition registrated by external $F^1$ , because
 for $F^2$ exists only unique proper time $\tau_2$ .
 The  new effect will be found only
when $F^1$ and $F^2$ will compare their initially synchronized clocks.

Analoguously to Classical Relativity average time boost depends on whether
 $F^1$ measures $F^2$ observables, as we considered or vice versa.
 To perform this measurement  we must have at least two synchronized
 objects with clocks $F^1_a$ and $F^1_b$ ,which make two $F^1$ and $F^2$ 
nonequivalent. 
If this experiment will be repeated several times
(to perform quantum ensemble) it'll reveal not only 
classical Lorentz  time boost ,
 but also the statistical spread having quantum origin and
proportional to $F^1$ time interval $\Delta t_1$ and $F^2$ momentum spread.

If the number of objects $N>2$
the modified clasterization formalism can be used ,
 which will be described here for the case 
 $N=3$  $\cite{Coe}$. According to previous arguments
 Hamiltonian in $F^1$  describing the two particles $G^2,G^3$ 
state evolution for  proper time $\tau_1$ is equal to
sum of two single particles energies. Rewritten through the
invariant system observables it have the form   :
\begin {equation}
   \hat{H^1}=
   m_1+[(s^{m}_{23})^2+\vec{p}^2_{23}]^\frac{1}{2} \label {C7}
\end {equation}
,where $s^m_{23}$ is  two particles $G^2, G^3$ 
invariant mass, which dependence on $\vec{q}_{23}$ is analogous to (\ref{C1}).
 In clasterization
formalism at first level we consider the relative motion of $G^2, G^3$
defined by
$\vec{q}_{23}$. At second level we regard them as the single quasiparticle
 - cluster $C_{23}$ with mass $s^m_{23}$ and $\vec{p}_{23}$  
  momentum in $F^1$. So at any level we regard 
 the relative motion of two objects only.
 Small Hilbert space $H_{23}$ 
 with the basis  $|\vec{q}_{23}\rangle$ can be extracted 
 from the total space $H^1_s$, which properties are the same
as for classic $F^1$ with infinite mass $\cite {Coe}$. 
This clasterization 
 procedure can be extended in the obvious inductive way to incorporate 
 an arbitrary number of the objects.

Due to appearance of quantum proper time 
 the transformation operator between two quantum RFs
$\hat{U}_{21}(\tau_2,\tau_1)$ is quite intricated, 
and to obtain it general form needs further studies. 
To illustrate the physical meaning of this transformation, we'll discuss
 briefly the transformation
  of single particle $G^3$ state between $F^1$ and $F^2$.
   We'll regard the particular case for which
  in $F^1$  at time $\tau_1^0=0$ the joint state vector
   of $F^2$ and $m_3$ - 
is $\psi^1_{in}(\vec{p}_{23},\vec{q}_{23})=\sum c^1_{jk}|\vec{p}_{23,j}\rangle|
\vec{q}_{23,k}\rangle$ is given, where $\vec{p}_{23}$ is $F^2,G^3$ total
momentum in $F^1$.  
   $F^2$ proper time $\tau_2$ is synchronized with $F^1$
   at this moment $\tau^0_2=\tau^0_1$.
  Due to unambiguous correspondence
 between the
   $\vec{p}_{13},\vec{q}_{13}$ and $\vec{p}_{23},\vec{q}_{23}$
phase space points we can assume that
   the state vector $\psi^2_{in}(\vec{p}_{13},\vec{q}_{13})$ in $F^2$
is obtained acting on $\psi^1_{in}$ by some unitary operator $\hat{V}_{21}$.
 Time evolution operator in $F^1$  $\hat{W}_1(\tau_1)=exp(i\tau_1\hat{H}^1)$
 is defined by Hamiltonian $\hat{H}^1$ only , and the same relation
is true for $\hat{W}_2(\tau_2)$ in $F^2$. Then $G^3$ state in $F^2$
at any $\tau_2$ can be obtained by the action of operator
$\hat{W}_2(\tau_2)\hat{V}_{21}\hat{W}_1^{-1}(\tau_1)$ on the
corresponding $G^3$ state in $F^1$. It means that despite $\tau_2$
and $\tau_1$ are correlated only statistically and have some quantum
fluctuations,  $G^3$ state vectors in 
$F^2, F^1$  at this moments are related unambiguously.


Now we'll consider obtained results in nonrelativistic limit.
It's easy to see that in the limit $\vec{p}_{12}\rightarrow 0$
 Hamiltonian (\ref{C2})
differs from $\hat{H_c}$ of (\ref{B5}) by the factor \\
$k_m=\frac{m_1+m_2}{m_1}$.
  It results from  energy-momentum Lorentz transformation from c.m.s. to $F^1$.
 We've chosen Newton-Wigner 
  space coordinate  operator in $F^1$ ${x}_{12}$ 
,which is canonical conjugated to $\vec{p}_{12}$.
In nonrelativistic limit it's equal to
$x_{21}=k_m^{-1}(x_2-x_1)$ , where $x_1,x_2$ are coordinates in ARF.
This result doesn't broke transformation invariance 
 , because there is no
established length scale in QM.
   
To illustrate our approach to quantum time
 we'll regard the simple model of the quantum clocks and RF
in which $F^1$ includes some ensemble (for example the crystal) of 
$\beta$-radioactive atoms $\cite{Per}$.
 Their nucleus can radiate neutrino $\nu$
(together with the electron partner)
which due to its very small cross-section practically can't be reflected
 by any mirror  and
reabsorbed by this nucleus to restore the initial state. Then for our purposes
this decay can be regarded as the irreversible stochastic process.
Taking the trace over $\nu$ degrees of freedom,
the final nucleus state can be described by the density matrix of
mixed state $\rho_N(t)$
and the proper time of this clocks of $F^1$ can be defined as:
$$ 
  \tau_1=-T_d\ln(1-\frac{N_d}{N_0})  
$$
where $N_0$ is the initial number of this atoms $N_d$ - the number of decays,
$T_d$ is the nucleus lifetime. The corrections to exponential decays,
appearing at time much larger then lifetime is neglected $\cite{Ghi}$.
It's easy to understand from the previous discussion how
 the superposition of Lorentz boosts can be applied to such
  system state, if its state vector has momentum spread .

  We consider in fact infrared 
limit for macroscopic object, so the role of negative energy states,which
is important for the standard relativistic problems must
be small. 

\section{\large{Concluding Remarks}}

  We've shown that the extrapolation of QM laws on free
macroscopic objects demands to change the approach to the 
space-time coordinate frames which was taken copiously from  Classical 
Physics. It seems that QM permits the existence of RF
 manifold, the transformations between which principally can't
be reduced to Galilean or Lorentz transformations.
It means that observer can't measure its own
spread in space, so as follows from Mach Principle it doesn't exist.
The physical meaning have only the spread of relative coordinates
of RF and some external object which can be measured by this RF or other
observer. 

Historically QM formulation started from defining the wave functions on
Euclidean 3-space $R^3$ which constitute Hilbert space $H_s$.
 In the alternative approach
accepted here we can regard $H_s$ as primordial states
 manifold. Introducing
particular Hamiltonian results in the relative asymmetry of $H_s$ vectors
 which permit
 to define $R^3$ as a spectrum of the continuous observable $\hat{\vec{r}}$
which eigenstates are
 $|\vec{r}_i>$. But as we've shown for several
 quantum objects one of which is RF
this definition
become ambiguous and have several alternative solutions defining $R^3$
on $H_s$. In the relativistic case the situation is more complicated, yet
as we've  shown it results in ambiguous Minkovsky space-time definition. 

In our work we demanded strictly that each RF must be quantum observer
i.e. to be able to measure state vector parameters. But we
should understand whether this ability is main property
 characterizing RF ? In classical Physics this ability 
doesn't influence the system principal dynamical properties. In QM at first
sight we can't
claim it true or false finally because we don't have the established theory
 of collapse. 
 But it can be seen from our analysis that collapse is needed
in any RF only to measure the wave functions parameters at some $t$.
Alternatively this parameters at any RF can be calculated given
the initial experimental conditions without performing the 
additional measurements.
It's quite reasonable to take that quantum states have objective meaning
and exist independently of 
their measurability by the particular observer,so this ability probably can't
 be decisive for this problem. It means that we can connect RF with the
system which doesn't include detectors ,which can weaken and simplify
our assumptions about RF. We can assume that primordial for  
 RF is the ability, which complex solid states have,
to reproduce and record the space and time points ordering  with
which objects wave functions are related. 

Author thanks M. Toller, A.Barvinsky, V. Bykov for fruitful discussions. 

\end{sloppypar}

\end{document}